# Charge-Order and Broken Rotational Symmetry in Magic Angle Twisted Bilayer Graphene

Yuhang Jiang[1], Xinyuan Lai[1], Kenji Watanabe[2], Takashi Taniguchi[2], Kristjan Haule[1], Jinhai Mao[1,3] and Eva Y. Andrei[1]

[1]Department of Physics and Astronomy, Rutgers University, New Jersey, 08854, USA

[2]National Institute for Materials Science, Namiki 1-1, Tsukuba, Ibaraki 305 0044, Japan

[3]School of Physical Sciences & CAS Center for Excellence in Topological Quantum Computation, University of Chinese Academy of Sciences, Beijing 100190，China

**The discovery of correlated electronic phases, including Mott-like insulators and superconductivity, in twisted bilayer graphene (TBLG) near the magic angle[1-4], and the intriguing similarity of their phenomenology to that of the high-temperature superconductors, has spurred a surge of research to uncover the underlying physical mechanism[5-9]. Local spectroscopy, which is capable of accessing the symmetry and spatial distribution of the spectral function, can provide essential clues towards unraveling this puzzle. Here we use scanning tunneling microscopy (STM) and spectroscopy (STS) in magic angle TBLG to visualize the local density of states (DOS) and charge distribution. Doping the sample to partially fill the flat band, where low temperature transport measurements revealed the emergence of correlated electronic phases, we find a pseudogap phase accompanied by a global stripe charge-order whose similarity to high-temperature superconductors[10-16] provides new evidence of a deeper link underlying the phenomenology of these systems.**

In the limit of strong correlations, the Coulomb interaction slows down the electrons or even localizes them in a Mott insulating phase, characterized by a spectral gap that opens at integer fillings. Doping a Mott insulator produces some of the most fascinating correlated quantum phases, such as the pseudogap and the high temperature superconductivity, with broken symmetries to accommodate the new order parameters. In the case of magic angle TBLG, doping the Mott-like insulating states leads to correlation driven superconducting phases at low temperatures[1,2]. Using STM/STS to study the electronic properties of magic-angle TBLG at temperatures above the superconducting transition, we observe a correlation driven pseudogap state characterized by partial gapping of the low-energy density of states (DOS). This pseudogap state is accompanied by a local charge polarization with a quadrupole symmetry and a global charge-ordered stripe phase which breaks the rotational symmetry of the moiré superstructure.

The STM topography of a TBLG near the magic angle measured in a gated device geometry (Fig. 1a) is shown in Fig.1b. The bright regions, which correspond to AA stacking where each top-layer atom is positioned directly above a bottom-layer atom, form a triangular moiré superlattice. The twist angle between the two layers, $\theta$, can be estimated from the moiré period, $L$, through the geometrical relation: $L \approx a/(2 \sin \theta /2)$, where $a$ = 0.246 nm is the graphene lattice constant[17,18]. Surrounding each bright region are six darker areas consisting of alternating AB /BA Bernal stackings. In the AB (BA) stacking each top-layer atom in the A (B) sublattice sits directly above a B (A) atom in the bottom layer, while top-layer B (A) atoms have no partner in the bottom layer. This is directly seen in the triangular structure of the atomic resolution topography, Extended Data Figure 1, where the triangles in the AB and BA regions show only one of the two sublattices. Earlier work on TBLG revealed two van Hove singularities (vHS) in the local DOS that flanked the Dirac point[17-21] with an energy separation of $\Delta E \sim \hbar v_F \Delta K_\theta - 2t_\perp$. Here $\hbar$ is the reduced Planck constant, $v_F$ the Fermi velocity, $\Delta K_\theta = 2K \sin(\theta/2)$, $K$ is the magnitude of the wave-vector at the Brillouin zone corner of the original graphene lattice, and $t_\perp$ is the interlayer hopping which can vary from sample to sample depending on the fabrication technique or external pressure[3]. As the twist angle is reduced, the two vHS approach each other until, at the "magic angle", $\theta_m \sim \frac{2t_\perp}{K\hbar v_F}$, they merge producing an almost flat band with vanishingly small Fermi velocity[19,22]. Due to the reduced kinetic energy, this narrow band is susceptible to interaction effects, so that when the Fermi energy ($E_F$) is brought within the band, interactions become dominant and the system can lower its energy through the formation of correlated electron phases.

We use STM topography together with STS measurements to identify the magic angle and the flat band condition in-situ. To study the electronic structure of TBLG near the magic angle we measure the dI/dV spectrum (I is the current, V the bias) which is proportional to the local DOS. The flat band in the DOS produces a peak in the dI/dV spectrum centered at the Dirac point (Fig. 1 c). The width of this peak is narrowest when two vHS have merged and the band is full or empty (Fig. 1c), providing a practical criterion for identifying the magic angle. Using the peak width as a guide, we find the narrowest peak in sample regions with no heterostrain[23], as identified by a perfectly triangular moiré pattern (Extended Data Figure 1). For the samples discussed here the magic angle is $\theta_m \sim 1.07°$ ($L \sim 13.2$ nm,), from which we extract $t_\perp \sim 104$ meV.

In order to separate the intrinsic band structure from correlation induced band-reconstruction effects, we first show (Fig.1c) the dI/dV spectra obtained in the AA region in the highly n-doped regime, for a gate voltage $V_g = 55$ V, where the band is fully occupied and in the highly p-doped regime ($V_g = -55$ V), where the band is empty. In both cases we observe a single spectral peak in the AA region, indicating the presence of the flat band. This peak is absent in the AB/BA regions (Fig. 1c inset), consistent with earlier reports[17,20,24]. The single peak structure is observed only very near the magic angle, but as the angle deviates from its magic value, the peak splits into two vHS that flank the Dirac point (Extended Data Figure 2), consistent with the single-particle band structure theory for TBLG[17,18]. It is worth noting that, while similar two vHS observed for non-magic twist angles are seen in all STS experiments, at the magic angle the DOS structure exhibits large sample to sample variations[25-27]. This reflects the fact that near the magic angle the band structure is very sensitive to the lattice relaxation, interlayer coupling, strain and interaction strength[23,28-30], which are influenced by sample preparation techniques and device geometry.

As we move the Fermi level into the flat band by adjusting the gate voltage, the single peak in the AA region broadens and splits revealing a pseudogap feature, Fig.1d, which suggests a band reconstruction associated with the emergence of a correlation-induced phase. In Fig. 1e we follow the evolution of the DOS in the center of the AA region with gate voltage, from highly p-doped where the band is empty ($V_g$ = -57 V) to the highly n-doped where it is full ($V_g$ = 69 V). For both the empty and the full bands the peak width at half height, ~ 40 mV, is narrowest and shows no evidence of splitting within the experimental resolution of ~1 mV. Once the Fermi level enters the flat band the peak begins to split. To compare our results with those obtained in transport measurements, we plot in Fig. 2a the doping dependence of the dI/dV intensity at the Fermi level. This procedure is adopted because it mimics the conductance measured with the transport technique, where only carriers near the Fermi level contribute. We note however that, due to their global nature, the transport measurements produce sample averaged global results, whereas the STS probes the local response. Despite the difference between these probes we find that, similar to the transport measurements, the doping dependence of the dI/dV near the Fermi level exhibits well defined dips near the same fillings, $\nu = 0, \pm 1/4, \pm 1/2, \pm 3/4, \pm 1$ , as those observed in the conductance measurements[1,2] corresponding to 4,3,2,1,0 electrons (+) or holes (-) per moiré cell, respectively.

Interestingly, as the Fermi level enters the flat band we observe a spectral weight redistribution away from the Fermi energy and into the sidebands[31]. This gives rise to a pseudogap feature flanked by two peaks whose relative height depends on doping as shown in Fig. 2b, 2c. We label the low energy and high energy peaks as lower band (LB) and upper band (UB) respectively, Fig. 2b. In Fig. 2c we zoom into the doping dependence near charge neutrality with 2V backgate intervals, corresponding to density steps of $1.4 \times 10^{11}$ cm$^{-2}$ or 0.2 electrons per moiré cell. At the charge neutrality point, $V_g$ ~ 0V, (yellow curves), the two peaks have roughly equal spectral weight and they are equidistant from the Fermi level with a separation of ~38 mV. Moving the Fermi level into the hole sector (green curves), we observe a spectral weight redistribution from the LB to the UB and vice versa when the Fermi level moves into the electron sector (red curves). This behavior closely resembles the spectral weight redistribution observed in earlier STS work on the doped Mott insulator phase in cuprates wherein, upon changing the doping level, the spectral weight transfers from one side of the Fermi level to the other[16], suggesting that the doping dependence observed here could be similarly attributed to a Mott-like insulating phase. The spectral weight transfer provides an independent method to monitor the local charge variation introduced by doping, and makes it possible to estimate the local charge from the ratio of the area under the dI/dV spectrum spanned by the LB, $A_{LB}$, to the total area spanned by the two peaks $A_{TOT}$. Here $A_{LB}$ is measured from the LB band edge to the Fermi energy after background subtraction, and $A_{TOT}$ is obtained by similarly measuring the area under the two peaks, Extended Data Figure 3. Using this procedure we obtain the local filling fraction directly from the spectral weight transfer: $v_A \sim 2\left(\frac{A_{LB}}{A_{TOT}} - \frac{1}{2}\right)$. For comparison we plot in Fig. 3e, the gate voltage dependence of $v_A$ together with the filling fraction values, $v$, calculated by the standard method of gate induced capacitive doping : $v = (V_g C L^2 \sqrt{3}/2)/4$ , where C is the capacitance between the TBLG, and 4 corresponds to the number of states in the hole or electron sector of the band (see Methods). The two curves closely track each other confirming that the spectral weight transfer provides a sensitive measure of the local charge. In what follows, we will use the spectral weight transfer as a local charge detector to identify the emergence of a correlated charge ordered phase.

Theoretically, the spectral weight distribution in a Mott insulating state can be qualitatively described by the Dynamical Mean Field Theory (DMFT)[32]. In Fig. 2d we plot the calculated local DOS for several filling fractions, projected to the AA-centered local functions, for the case of the fully filled band as well as for several fractional fillings of the band as obtained by the DMFT simulation (see Methods). In the extreme n-doped limit, corresponding to the full band ($v = 1$), DMFT predicts a single peak similar to the experimental result, even though the intrinsic non-interacting DOS is split into two narrow peaks, because the Coulomb interaction broadens the peaks and smears out the double-peak structure. At fractional fillings ($v = 0, \pm 1/4, \pm 1/2$), DMFT opens a Mott gap in the local spectra, in which the occupied (empty) bands correspond to the lower (upper) Hubbard bands. The lower Hubbard band loses spectral weight with doping, while the upper band gains the weight, in qualitative agreement with the STM measurement.

To reveal the electronic structure, and the underlying symmetry of the emergent correlated state in the partially doped flat band, we study the spatial dependence of the topography, Fig. 3a, and spectroscopy, Fig. 3b-d, at the charge neutrality point, $V_g = 0$ V. In topography, the AA (bright) region shows up as a perfect disk, consistent with the symmetry of the moiré pattern which, in the limit of small twist angles and a continuum approximation[8], is $C_6$. In contrast to the topography, the dI/dV maps taken in the same region at bias voltages corresponding to the LB (Fig. 3b left panel) and the UB (Fig. 3b right panel) form ellipses with major axis orientations that are roughly orthogonal to each other and clearly break the $C_6$ symmetry. To elucidate the origin of the broken symmetry revealed by the dI/dV maps, we plot in Fig. 3d the position dependence of the spectra taken along the arrow shown in Fig. 3a. We note that, even though the backgate voltage is fixed at $V_g = 0$ V, we nevertheless observe a clear spectral weight transfer from the LB which is dominant in the bottom part of the AA region (red end of the arrow) to the UB which dominates in the top part of the AA region (green end of the arrow). This is remarkably similar to the spectral weight transfer observed in the center of the AA region as a function of doping (Fig. 2c). Since the backgate voltage is set to a fixed value ($V_g = 0$ V), it is then natural to interpret the position dependence of the spectral weight transfer in terms of a spatial charge redistribution within the AA region. Using the same analysis as in Fig. 3e to calculate the local charge from the peak areas $A_{LB}$ and $A_{TOT}$ we obtain the position dependence of the local charge shown in Fig. 3f. Following this analysis, the high intensity elliptical region for the UB map corresponds to hole doping whereas the high intensity elliptical region for the LB map corresponds to electron doping. In order to extract the spatial dependence of the local doping in the flat band, we subtract the intensity of the UB peak from the LB peak at every point on the map in Fig. 3b. Following this subtraction procedure (Methods), which directly visualizes the spatial charge modulation (Fig. 3c), we observe that within the AA region (red circle) the charge modulation exhibits a quadrupole geometry, consisting of four quadrants with alternating positive (blue) and negative (red) charge, ranging from ~1 electron to ~1 hole per moiré cell. Similar charge polarization patterns are observed for $\nu = \pm 1/4$ fillings (Extended Data Figure 4). Importantly, as the Fermi level is outside the flat band (Extended Data Figure 5), or for twist angles far from the magic value (Extended Data, Figure 6), the $C_6$ symmetry of the spectral maps is restored. This confirms that the broken symmetry is associated with a correlation induced ordered phase, and rules out interpretations in terms of artifacts such as tip anisotropy[33], or strain effects (see Methods).

In addition to the local charge ordering observed in the scans of Fig. 3, large area scans (Fig. 4) exhibit globally ordered charge-stripes aligned with a crystallographic axis of the moiré superlattice, which break the initial $C_6$ symmetry of the moiré lattice reducing it to $C_2$. As seen in Fig. 4 c, the global stripe order emerges from the alignment of the quadrupole charge lobes that are centered on the AA regions. The individual quadrupole lobes are not perfectly aligned with the stripe direction, but are slightly tilted (~ 16°) with respect to it (Extended Data Figure 7). While we observed the charge symmetry breaking in all magic angle samples studied here, (Extended Data Figure 8,9), we expect that there are many competing quantum states near the magic angle, and therefore slightly different sample preparations could possibly result in different correlated ground states.

Earlier reports have shown that the phenomenology of magic angle TBLG bears a tantalizing resemblance to that of high temperature superconductors[1,2]: the emergence of superconductivity in a doped Mott-insulator, the dome-like doping dependence of the superconducting critical temperature, the linear temperature dependence of the resistivity, and the relatively large ratio of critical temperature to the Fermi energy. Both the pseudogap phase and its stripe charge-order revealed here, closely resemble similar observations in the pnictides[13,14] and cuprates[12,15], where the high temperature superconductivity is believed to be intimately connected to the pseudogap phase and the stripe order is common[10]. The findings reported here add a new piece to the puzzle connecting TBLG and high temperature superconductors, and identify an important constraint towards unraveling the nature of the correlated electronic states in these systems.

**Acknowledgments**

We acknowledge support from NSF-DMR 1708158 (Y.J.), DOE-FG02-99ER45742 (E.Y.A and J.M), National Key R&D Program of China (Grant no. 2018YFA0305800) (J.M.), NSF-DMR 1709229 (K.H.). We thank R. Fernandes and Z. Bi for stimulating discussions.

**Main Figure Legends:**

**Figure 1. STM/STS near the magic angle. a**, Schematic of sample geometry and STM experiment. **b**, STM topography of TBLG (upper pannel) for twist-angle $\theta \sim 1.07°$ ($V_b = 200$ mV, $I = 10$ pA). The different stacking configurations (AA, AB and BA) discussed in the text are shown in the lower panel. **c,** dI/dV spectra taken in the AA regions for the fully occupied ($V_g = +55$ V) and empty ($V_g = -55$ V) band . ($V_b = -200$ mV, $I = 20$ pA), The Fermi energy for all STS spectra defines the energy origin. The spectrum in the AB region taken at $V_g = +55$ V is shown in the inset. **d**, dI/dV spectrum in the center of the AA region, at the charge neutrality point showing the emergence of a pseudogap that splits the DOS peak. $V_b = -200$ mV, $I = 20$ pA. **e,** evolution of the peak with gate voltage in the center of the AA region, from highly p-doped where the band is empty ($V_g = -57$ V) to the highly n-doped where it is full ($V_g = 69$ V).

**Figure 2. Doping dependence of dI/dV spectra. a**, Gate voltage (filling fraction) dependence of the dI/dV intensity at the Fermi level shows clear dips at fillings fractions of $\nu = 0, \pm 1/4, \pm 1/2, \pm 3/4, \pm 1$. Here + (-) correspond to the electron (hole) doped sectors respectively. **b**, Color map of gate voltage dependent dI/dV spectra in the AA region highlighting the doping induced spectral shift between the LB and UB. The dashed line marks the charge neutrality point. $V_b = -300$mV, $I = 20$pA. **c**, Gate voltage dependence of dI/dV spectra taken at the center of an AA domain close to charge neutrality. The curves (offset for clarity), correspond to doping levels ranging from $-0.86 \times 10^{12}$ cm$^{-2}$ ($V_g = -12$ V) to $0.86 \times 10^{12}$ cm$^{-2}$ ($V_g = +12$ V). The spectral weight shift between the LB and UB with doping is clearly seen. **d**, DMFT simulation of local DOS projected to the AA-centered local functions at different filling fractions as discussed in the text.

**Figure 3. Spatial charge modulation in the correlated phase. a**, STM topography in a 10x10nm$^2$ area centered on an AA region ($V_b$ = -200 mV, $I$ = 20 pA). The red circle labels the AA region. **b**, dI/dV map over the same area as panel **a** for the LB (left panel) and UB (right panel) at $V_g$ = 0 V (10 × 10nm$^2$, $V_b$ = -200 mV, $I$ = 50 pA). **c**, Map of the net charge obtained by the method described in the text. Red corresponds to electron doping and blue to hole doping. The four dashed lobes mark sectors of alternating electron (e) and hole (h) doping. **d**, Spatial dependence of dI/dV curves along the colored arrow in **a** and **c** shows the spectral weight shift between the LB and UB with position. **e,** Gate voltage dependence of filling fraction (symbols) within the flat band extracted from Fig. 2**c**. The dotted line shows the gate dependence of the filling fraction, ν, obtained from the gate voltage as described in the text. **f,** Position dependence of filling fraction from **d** along the path indicated by the arrow in **c**. The filling fraction was obtained from the normalized area under the LB peak as discussed in the text.

**Figure 4. Global stripe charge-order**. **a-b**, Large area dI/dV map for the LB (-29 mV) (**a**) and UB (22 mV) (**b**) in the AA areas at charge neutrality ($V_g$ = 0 V) ($V_b$ = -300 mV, $I$ = 70 pA). **c**, Map of net charge obtained by the method described in the text and in Fig. 3**c**. Red corresponds to electron doping and blue to hole doping. The four lobes superposed on each AA area mark the alternating sectors of electron (red) and hole (blue) doping.

# Figure 1

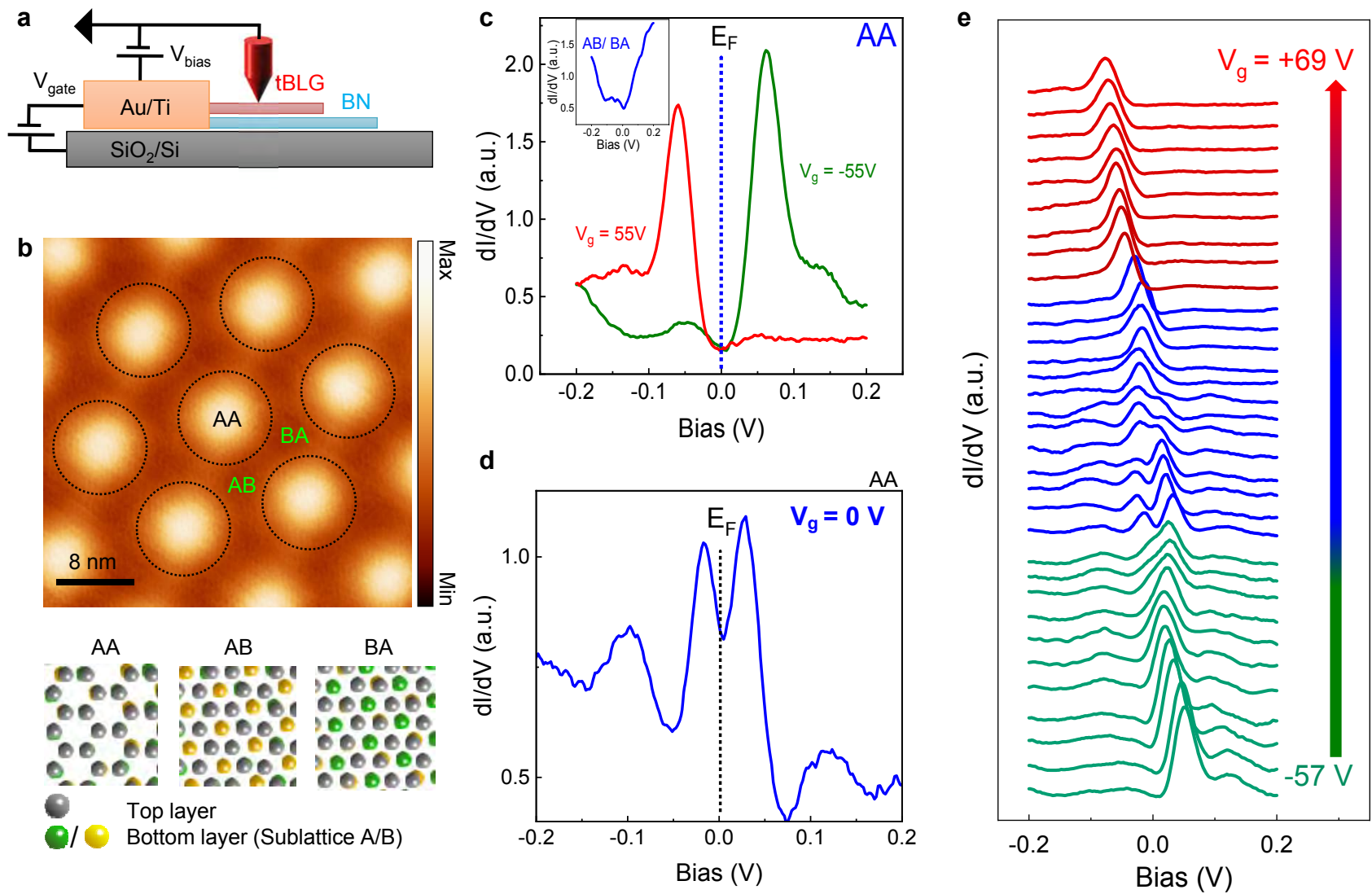

Figure 2

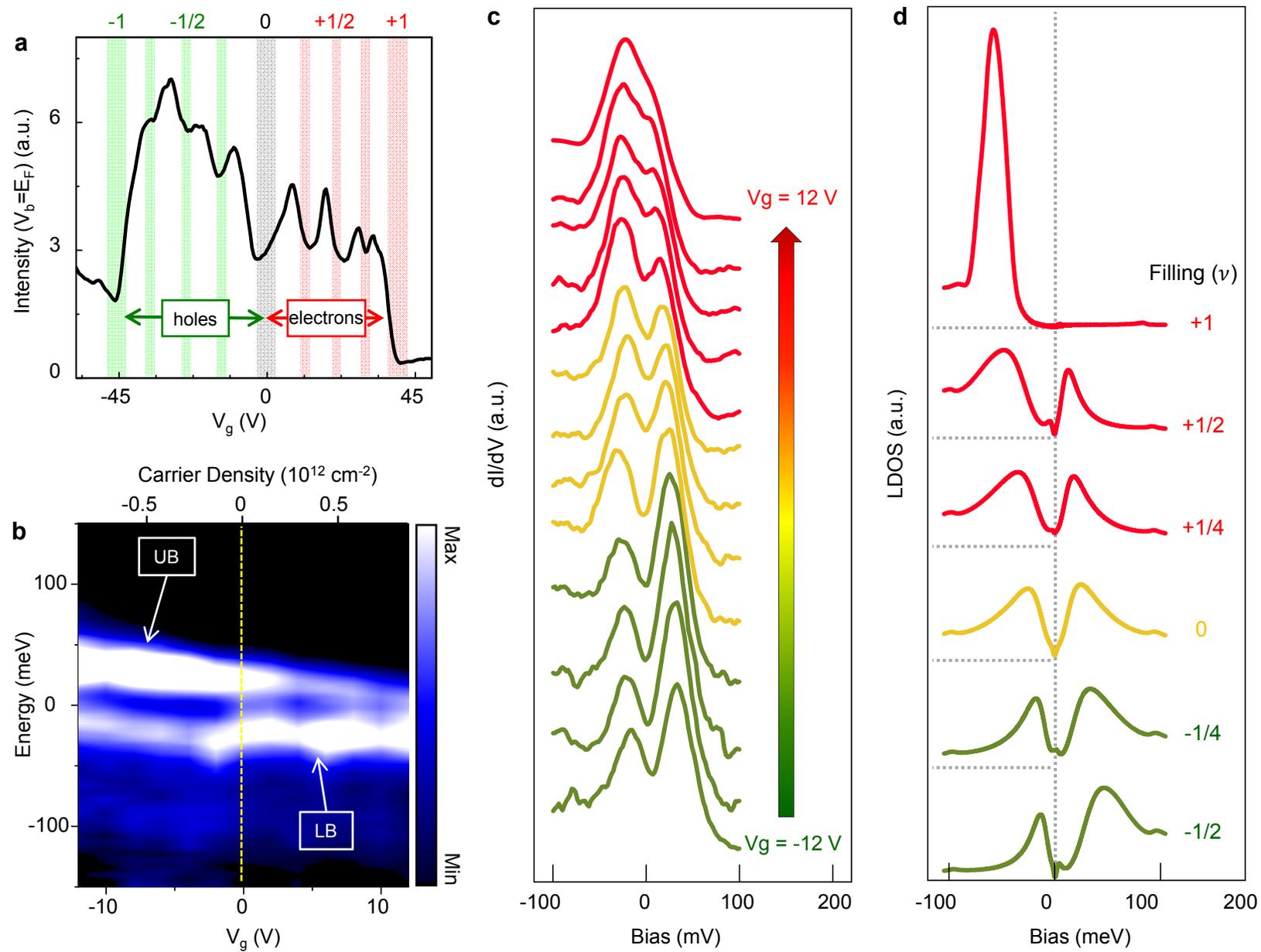

# Figure 3

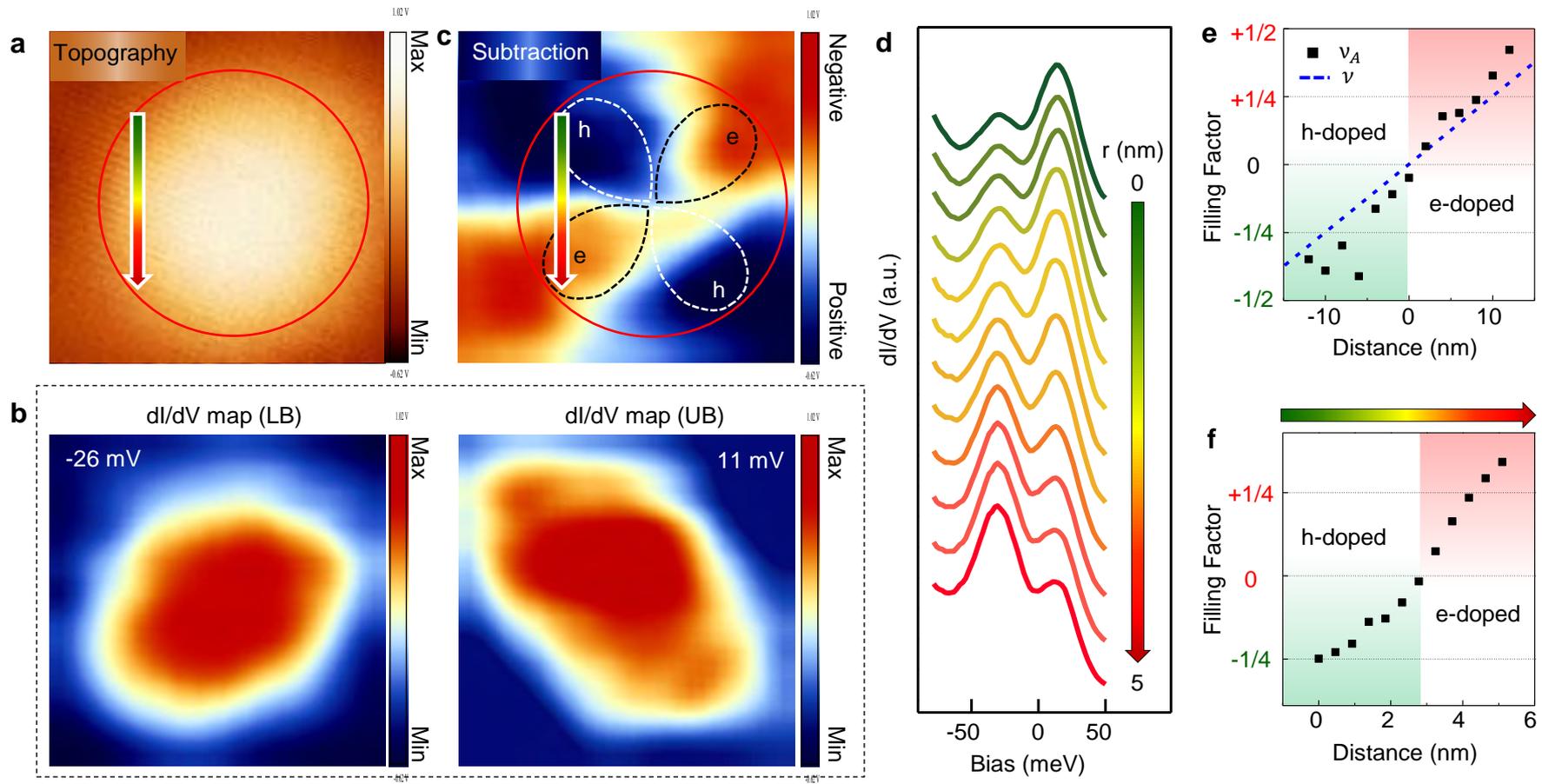

Figure 4

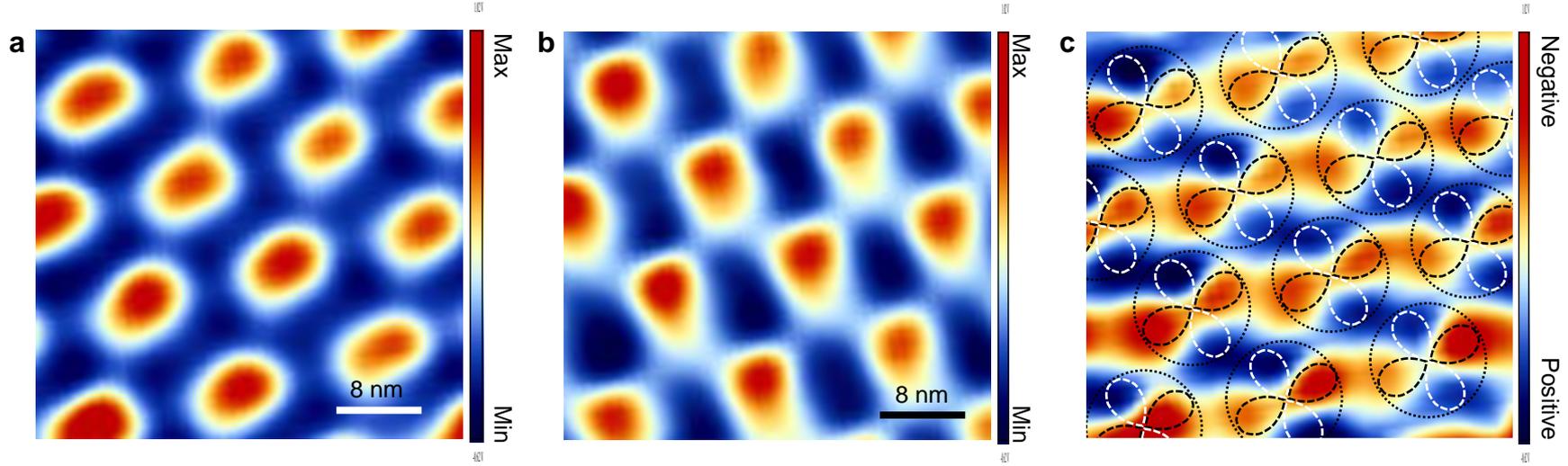

**Methods**

**Sample Preparation.** The samples are fabricated by the tear and stack method[34]. A PVA (poly vinyl alcohol ) was spin coated on the silicon chip and baked at 80 °C for 10mins, followed by PMMA (Poly methyl-methacrylate) spin-coating and another round of baking at 80 °C for 30 mins. Graphene was next exfoliated on top of the PMMA/PVA film. The PMMA thin film was peeled off from the PVA by the tape with pre-cut window for subsequent transfer on an hBN flake which was pre-exfoliated on $SiO_2$ and annealed in forming gas (10% $H_2$, 90%Ar) at 300 °C for 2 hours for better adhesion. Next, part of the graphene flake was brought in contact with the hBN surface forming a G/hBN/$SiO_2$ heterostructure. The other part of the flake which adhered to the PMMA was detached upon lifting the film. Subsequently the G/hBN/$SiO_2$ structure was rotated by 1.1° ± 0.2° before the next transfer step where the second part of the graphene flake was deposited on the heterostructure, forming the TBLG with the designated twist-angle, and the PMMA film was removed. The success of this procedure relies on the higher adhesion between G and hBN compared to G on PMMA. In the final fabrication step a Ti/Au electrode was evaporated for the electrical contact through a shadow mask. All the samples are annealed in forming gas (10% $H_2$, 90%Ar) at 260 °C overnight to remove all fabrication residues and allowfor lattice relaxation.

The STM experiment was performed in a home-built setup at 4.6 K using a chemically etched tungsten tip. The dI/dV spectra were collected by the standard lock-in technique, with 1 mV A.C. voltage modulation at 617.3 Hz added to the D.C. bias[35,36]. To simplify the discussion in the main text, we define the $V_g = V_{BG} - V_{G0}$, where the $V_{BG}$ is the applied backgate voltage, $V_{G0}$ = -7 V is the backgate voltage corresponding to the charge neutrality point.

**dI/dV spectrum evolution of vHS away from magic angle.** In the main text, we have shown the gate dependent behavior for TBLG at 1.07°, where partial filling of the flat band led to a splitting of the DOS peak, signaling the emergence of a correlation induced pseudogap phase accompanied by stripe charge order. For twist angles away from the magic angle doped far from charge neutrality, two vHS are observed instead of the single peak at 1.07°, as shown in Extended Data Figure 2 (a), (b) where the twist angle is

$\theta \sim 1.2°$. When the Fermi level is tuned into such a vHS peak, we can still observe splitting as long as the twist angle is close to the magic angle. However, at larger angles no splitting is observed when the Fermi level is brought into the peak, as shown in Extended Data Figure 2 (c) where $\theta \sim 1.7°$.

**Estimating the filling fraction.** As discussed in the main text, the area under the dI/dV spectrum can be utilized as an independent measure of the filling fraction. To illustrate the procedure we consider in Extended Data Figure 3 a spectrum at filling $\nu \sim -0.3$ (left panel) as calculated from the value of $V_g = -10$ V. To extract the area under the spectrum, we first subtract the background shown by the blue dashed line, which runs through the lower and upper edges of the band. $A_{LB}$, marked in red, which corresponds to the LB, is calculated by integrating the area under the curve from the lower edge of the band to the Fermi energy. From the ratio between $A_{LB}$ and the total area, $A_{TOT} = A_{LB} + A_{UB}$, we calculate the filling fraction: $\nu_A \sim \left(\frac{A_{LB}}{A_{TOT}} - \frac{1}{2}\right) \times 2 \sim 0.39$. We find that $\nu_A$ is comparable to the value, $\nu$, extracted from the backgate voltage.

**Calculation the filling fraction from the gate voltage needed to span the entire band.** In Fig. 2a of the main text, we studied the evolution of the dI/dV spectra with the backgate voltage, and showed the dI/dV ($E_F$) intensity as a function of filling. In our data, the band becomes fully filled (8 electrons per moiré cell) for $V_g \sim +40$ V and it is emptied for $V_g \sim -40$ V. By considering the ~300 nm thick silicon oxide, the 20nm hBN substrate and the triangular moiré pattern with wavelength about 13.2 nm, we find that 80 V of gate voltage delivers ~ 8 electrons per moiré cell, in agreement with the expected value.

**Broken $C_6$ symmetry in magic angle TBLG at different fillings.** In Fig. 3, 4 of the main text, we observed the charge ordered phase through the spatial dependence of the spectral weight transfer between the LB and UB in the globally neutral state, $V_g = 0$ V. Similar analysis shows that the electronic ordering is not limited to the charge neutrality point but can be observed at other fillings (±1/4) within the flat band as shown in Extended Data Figure 4.

**Absence of broken symmetry in the full band.** We have shown the broken $C_6$ symmetry and charge modulation occurs in the partially filled flat band. To address the question whether this broken symmetry is a property of enhanced interactions in the partially filled flat band or of the band structure itself, we show in Extended Data Figure 5 the topography and dI/dV map for the full band ($V_g$ = +55 V) taken over the same area as Fig. 4 in the main text. The dI/dV maps taken at a bias voltage corresponding to the center of the flat band (dashed line in the inset of b) show circular AA areas as opposed to the elliptical shapes observed in the partially filled band, indicating a fully recovered $C_6$ symmetry.

**Absence of broken symmetry at non-magic twist-angles.** In Extended Data Figure 6 (a), we show an area where the twist-angle angle ($\theta$ = 1.5°) is away from the magic angle. The moiré period here is ~ 9.4 nm, and the DOS shows two vHS peaks separated by 80 mV (insert). The dI/dV map at the energy of one of the vHS, Extended Data Figure 6 (b), where the circular shape of the AA regions (dashed circle), indicates that the rotational symmetry is preserved away from the magic angle.

**Different charge-order orientations.** Extended Data Figure 8 shows a different area (several micrometers away, which means the sample is moved by the coarse motor and the tip is landed again) of magic angle TBLG where the stripe orientation is slightly different. This indicates the presence of charge order domains. Importantly the observation of domains with different stripe orientations helps rule out interpretations in terms of artifacts such as tip anisotropy.

**Charge-order in different sample.** The charge ordered phase that forms in the partially filled flat band in the magic-angle TBLG was observed repeatedly both in different regions of the same sample as well as in other samples where the twist between the layers corresponded to the magic angle. Extended Data Figure 9 shows the charge modulation in the AA region near the $\nu \sim 0$ filling in a different sample. Both the charge modulation (Extended Data Figure 9**c**) and the one dimensional stripe charge-order (Extended Data Figure 9**f**) are observed in this sample.

**Excluding possible artifacts.** Here we use the following results to exclude possible artifact-induced charge order effect observed in this work. (1) With the same tip in the same area, we observe no broken symmetry in the full band as shown in Extended Data Figure 5. This excludes the possibility that the observed charge-order is a result of local strain or an artifact arising from an anisotropic tip. (2) We observe no broken symmetry at non-magic twist-angles (Extended Data Figure 6) again ruling out artifacts such as strain or an anisotropic tip. (3) The energy shift in the LB and UB as a function of position can also help to rule out the tip effect. In Extended Data Figure 10, we show an extension of the data in Fig. 3d taken out to a distance of 6.8 nm along the colored arrow. On this length scale an energy shift in both LB and UB is clearly seen. (4) In extended data Figure 9, the shape of the high intensity regions of the DOS maps measured in the LB and UB are more complex than those in Fig. 3b. While a distorted shape could be attributed to tip anisotropy, this explanation is inconsistent with the fact that the shape changes with bias voltage, with position in the sample, and with the fact the energies of the UB and LB change with position along the sample.

**DMFT simulations for doping dependent LDOS for flat band.** To construct the tight-binding model, we followed Refs. 19,37 and approximate the Fourier transform of the interlayer tunneling by $t(q)=t_0 \exp(-\alpha (q\,d)^{\gamma})$ with parameters[19,37] ($t_0 =1.066$ eV, $\alpha = 0.13$, $\gamma = 1.25$, $d = 3.34$ Å). The hopping within the graphene layer is $t_{gr} = 2.73$ eV. For the DMFT calculation, the set of four localized correlated orbitals is derived in Ref. 30. While these orbitals are very localized and contain almost all the weight of the narrow band, their projection to the tight-binding bands breaks the particle-hole symmetry, so that the occupied (unoccupied) part of the higher-energy band at the Gamma point has large (small) projection to the localized set of orbitals. Consequently, the DMFT correlated spectra appears asymmetric at the charge-neutral point ($\nu = 0$), in disagreement with experiment. In the orbital set of Ref. 30 we use equal phase for the projection to the top and the bottom layer of the bilayer graphene. We found that the phase-difference between the localized orbitals on the top and the bottom layer controls this particle-hole asymmetry. Moreover, we found that the purely imaginary phase between the two layers forms an optimized projector for DMFT method, which not only restores the particle-hole symmetry, but also increases the amount of overlap between the narrow band and the localized set of functions (from 95% to 99.9%). Hence, this modified set of localized orbitals was utilized here. Finally, the Coulomb interaction among the correlated orbitals is accounted for by the DMFT method, where we assumed the screening of the interaction by dielectric constant $\varepsilon = 5$, which leads to a Coulomb interaction strength of $U = 150$ meV on the local orbitals, as in Ref 30.

**Data Availability**

The data that support the findings of this study are available from the corresponding author on reasonable request.

**Extended Data Figure Legends**

**Extended Data Figure 1 Atomic resolution of AB and BA regions near the magic angle. a**, Large area STM topography 8.9 nm × 8.9 nm, $V_b = -200$ mV, $I = 30$ pA. **b-c,** zoom in of the area in (**a**) as indicated by the green and blue square. Schematic drawing of graphene lattice is superposed on the STM topography to highlight the sublattice polarization due to the AB (BA) stacking. (**d**) Topography of undistorted moiré pattern at for TBLG at the magic angle indicating the absence of heterostrain.

**Extended Data Figure 2 dI/dV spectra and evolution of vHS with doping away from the magic angle.**
**a,** dI/dV spectra for TBLG at $\theta \sim 1.2°$. **b,** Zoomed-in image of the black dashed rectangle in the (**a**). **c,** Evolution of the dI/dV spectra with doping for a TBLG at $\theta \sim 1.7°$.

**Extended Data Figure 3 Estimate of the filling fraction from the area under the LDOS peak. a,** dI/dV curve for $\nu \sim -0.3$. The dashed line represents the background subtraction. **b,** dI/dV spectrum after background subtraction. Colored areas are used to estimate the filling fraction as described in the text.

**Extended Data Figure 4 Charge polarization within moiré cells of TBLG at ±1/4 filling.** dI/dV curves and maps at +1/4 filling (**a**) and -1/4 filling (**b**) ($V_b$ = -200 mV, $I$ = 50 pA). The left panels show the dI/dV curves at ±1/4, the center panels show the dI/dV maps at the LB energy, and the right panels show the dI/dV maps at the UB energy.

**Extended Data Figure 5 Absence of broken symmetry in the full flat band.** STM topography (**a**) and dI/dV map (**b**) in same area measured at the energy corresponding to the center of the flat band in the highly n-doped regime ($V_g$ = +55 V) corresponding to the fully filled flat band ($V_b$ = 200 mV, $I$ = 15 pA).

**Extended Data Figure 6 Absence of broken symmetry at non-magic twist-angles. a,** STM topography of tBLG away from magic angle ($\theta$ = 1.5°) centered on the AA region ($V_b$ = -150 mV, $I$ = 20 pA). The dashed circle labels the AA region. Insert: dI/dV spectrum in AA regime ($V_b$ = -150 mV, $I$ = 50 pA). **b,** dI/dV map over the same area as panel (**a**) at the energy of left vHS (-29 mV) labled by the dashed line in the insert in (**a**) ($V_b$ = -150 mV, $I$ = 50 pA).

**Extended Data Figure 7 Relative orientation between the negative charge lobes and the charge stripe direction. a,** Charge order extracted from large area showing stripe charge orientation along a crystallographic axis of the moiré lattice (same as Fig. 4c). **b,** The relative orientation between the charge quadrupole lobes (green lines) and the charge stripes (red lines), $\theta \approx 16 \pm 2°$, is roughly constant within this region.

**Extended Data Figure 8 Different charge-order orientations. a,** dI/dV maps at the energy of LB (left panel) and UB (right panel) for the same sample as that discussed in the main text, but in a different region where the charge stripe is along a different direction. **b,** Charge modulation map obtained by subtracting the two intensity maps as discussed in the main text. The black arrow labels the direction of the electron lobe and the green arrow marks the direction of the global charge stripe which coincides with a crystallographic axis of the moiré pattern. **c,** For comparison we show Fig. 3c in that main text where the orientation labeled by the arrow has changed.

**Extended Data Figure 9 Charge modulation in another TBLG sample near the magic angle. a,** STM topography in a 10 nm × 10 nm area centered on an the AA region ($V_b$ = -100 mV, $I$ = 40 pA). The red circle labels the AA region. **b,** dI/dV map over the same area as panel (**a**) for the LB (left panel) and UB (right panel) at $V_g$ = 0V ($V_b$ = -100 mV, $I$ = 40 pA). **c,** Map of net charge obtained by the method described in the main text. Red corresponds to electron doping and blue to hole doping. The four dashed lobes mark the sectors with alternating electron (e) and hole (h) doping. **d,** Spatial dependence of dI/dV curves along the colored arrows in **a** and **c** shows the spectral weight shift between the LB and UB with position. **e,** Position dependence of filling fraction from **d** along the path indicated by the arrow in **c**. The filling fraction was obtained from the relative area under the LB peak as discussed in the text. **f,** Large-scale dI/dV map (40 nm) of net charge obtained by the method described in the text and in Fig. 3**c**.

**Extended Data Figure 10 Energy shift in the LB and UB peak positions correlated with the C2 pattern.** In the left panel (Fig. 3d), that the distance, $r$ ~5 nm, is too small to distinguish the energy shift in two bands (dashed lines). So we extend the length to 6.8 nm shown in the right panel, where an energy shift in both LB and UB can be seen.

.

# Extended Data Figure 1

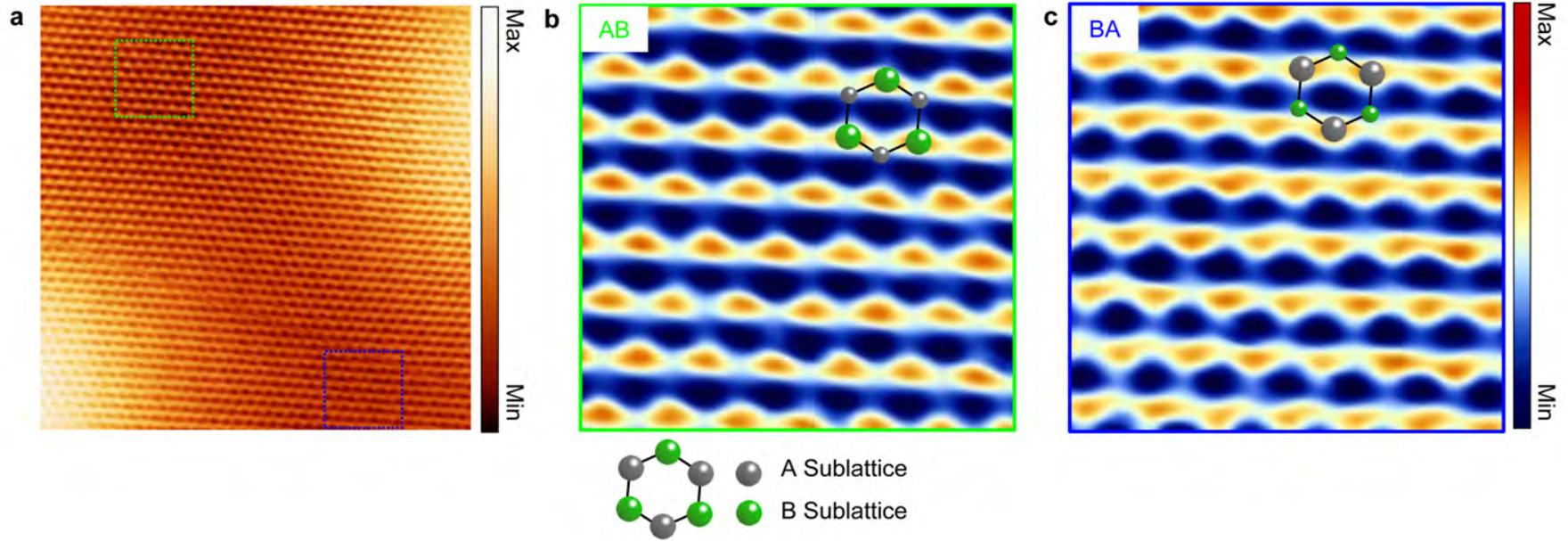

# Extended Data Figure 2

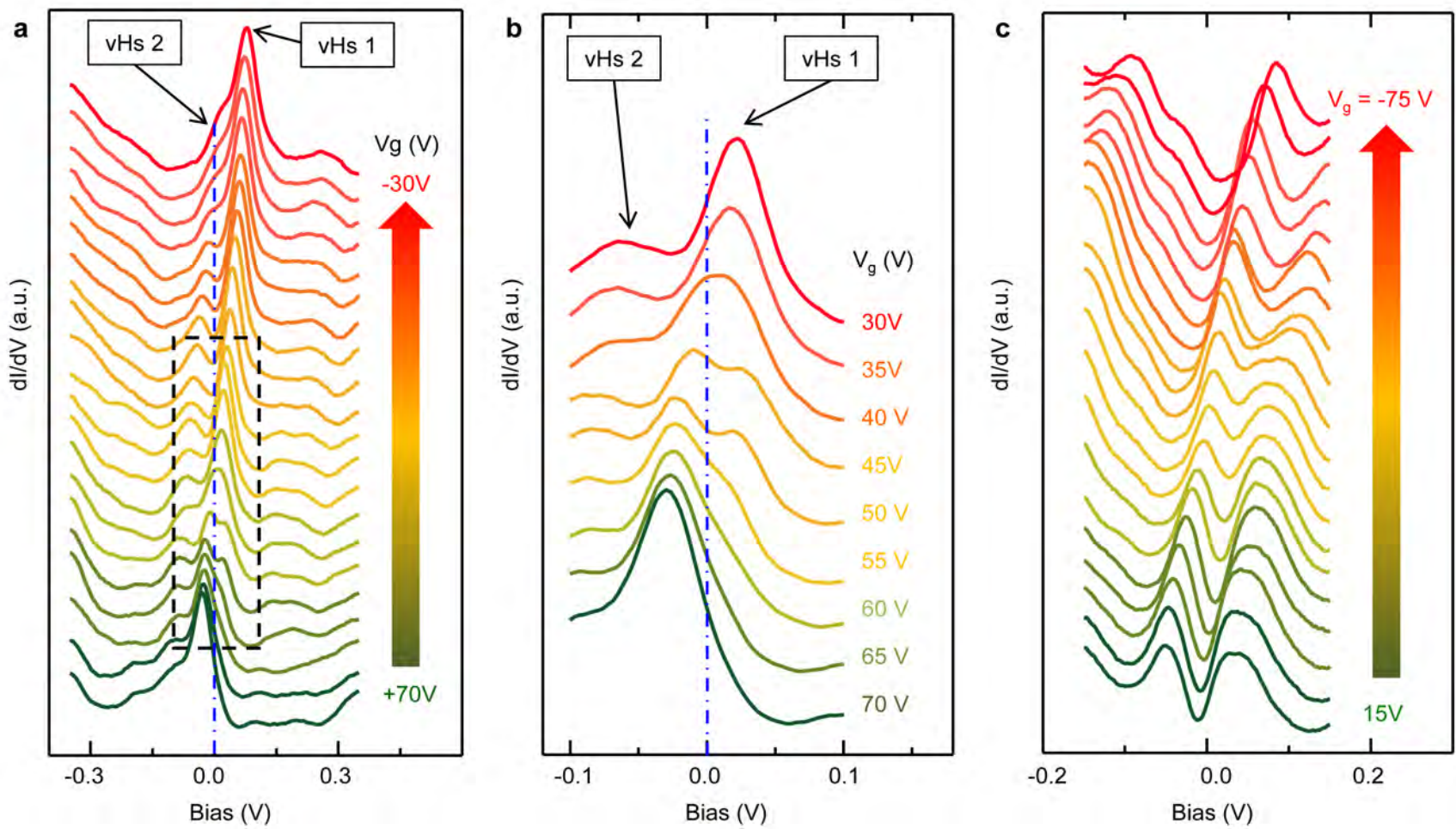

# Extended Data Figure 3

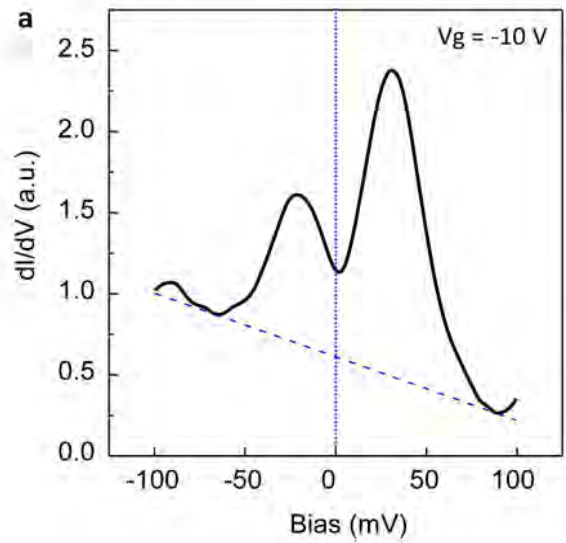 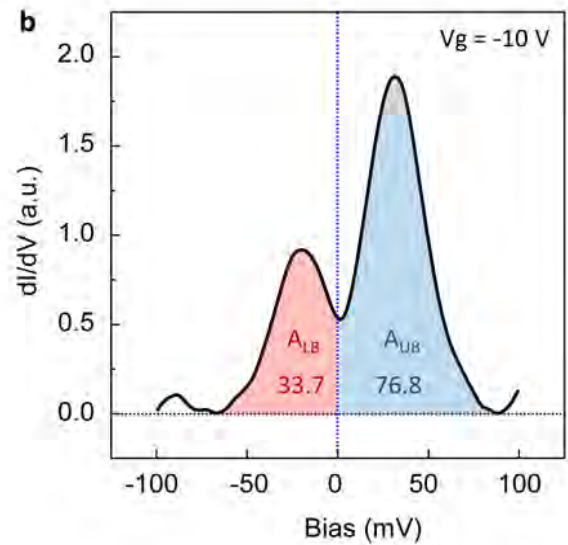

# Extended Data Figure 4

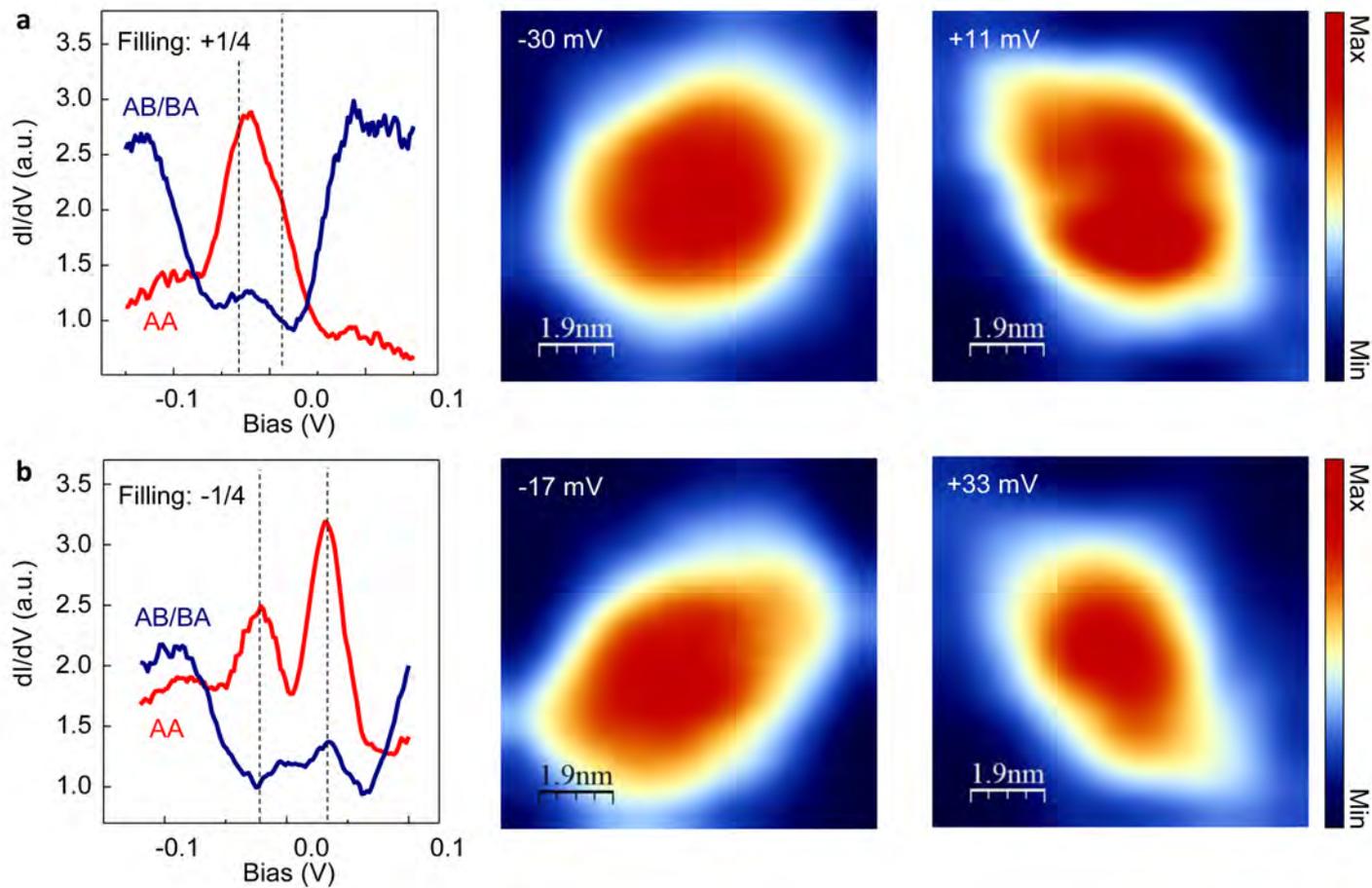

# Extended Data Figure 5

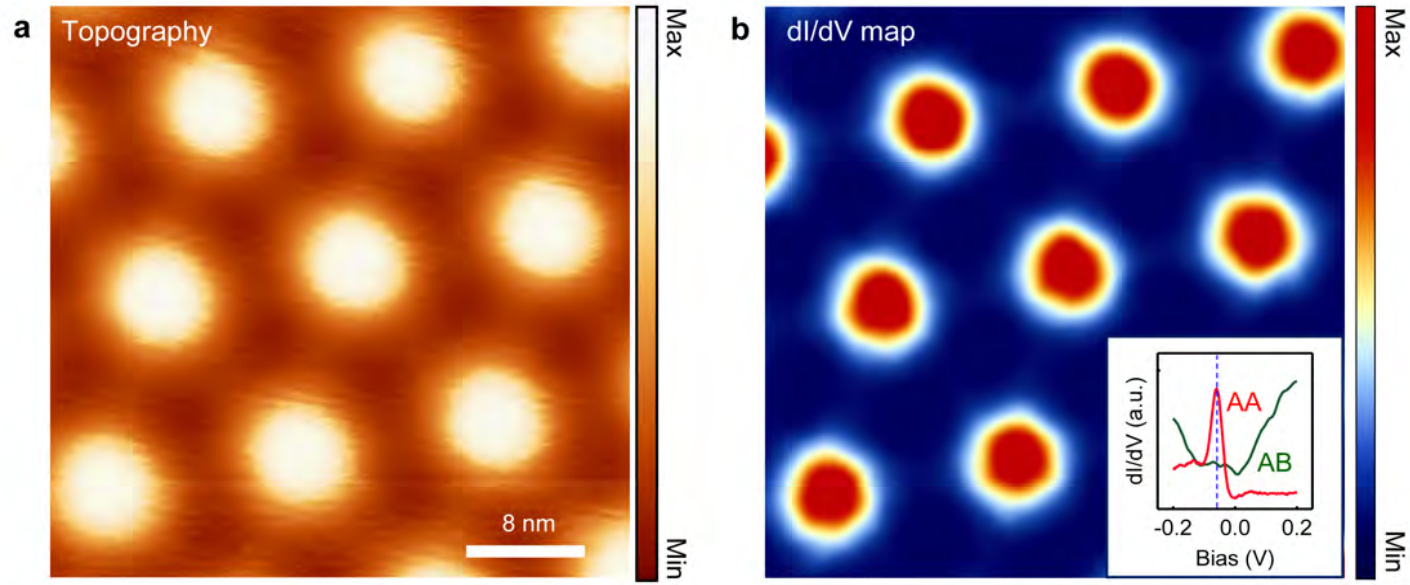

# Extended Data Figure 6

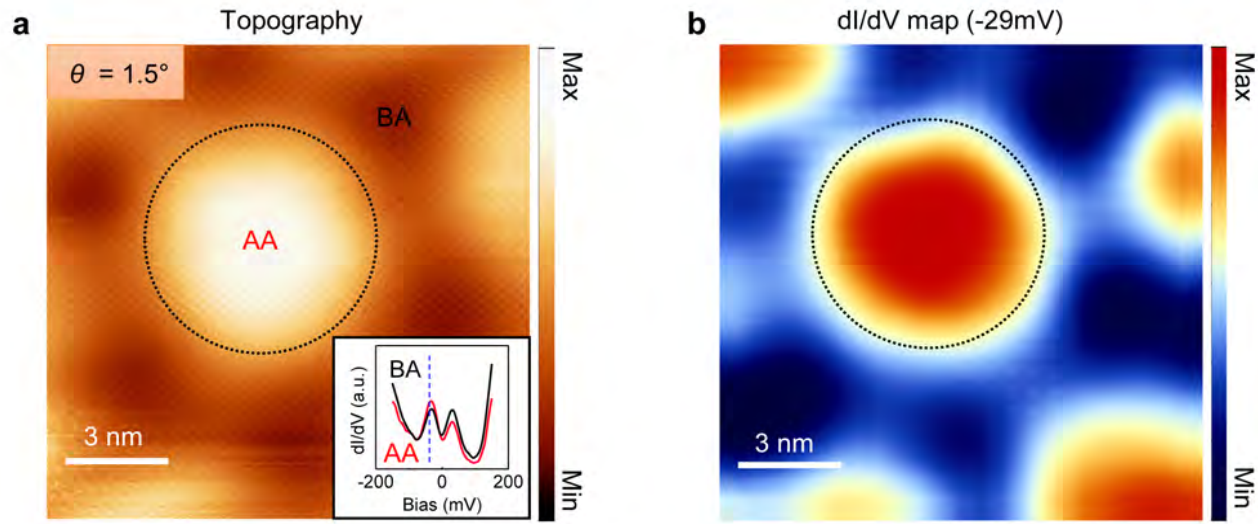

# Extended Data Figure 7

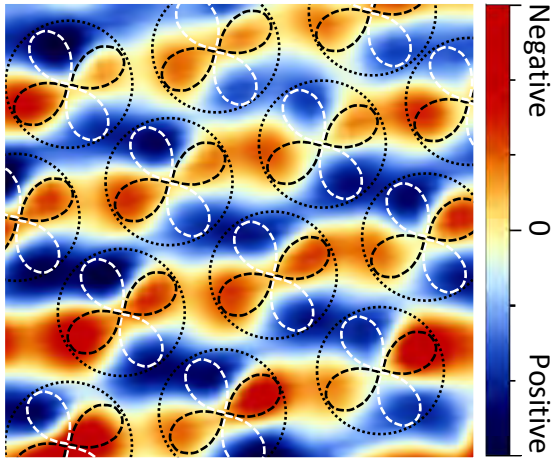 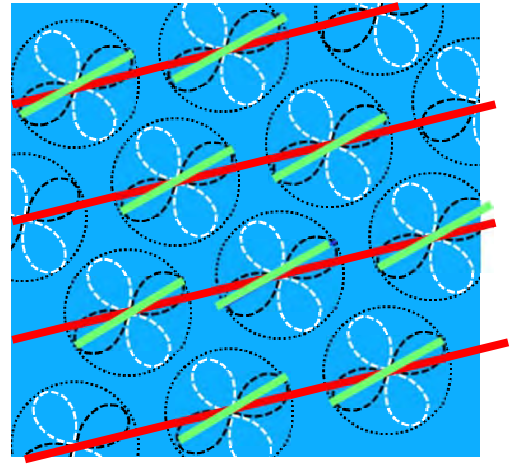

$\theta = 16^0 \pm 2^0$

# Extended Data Figure 8

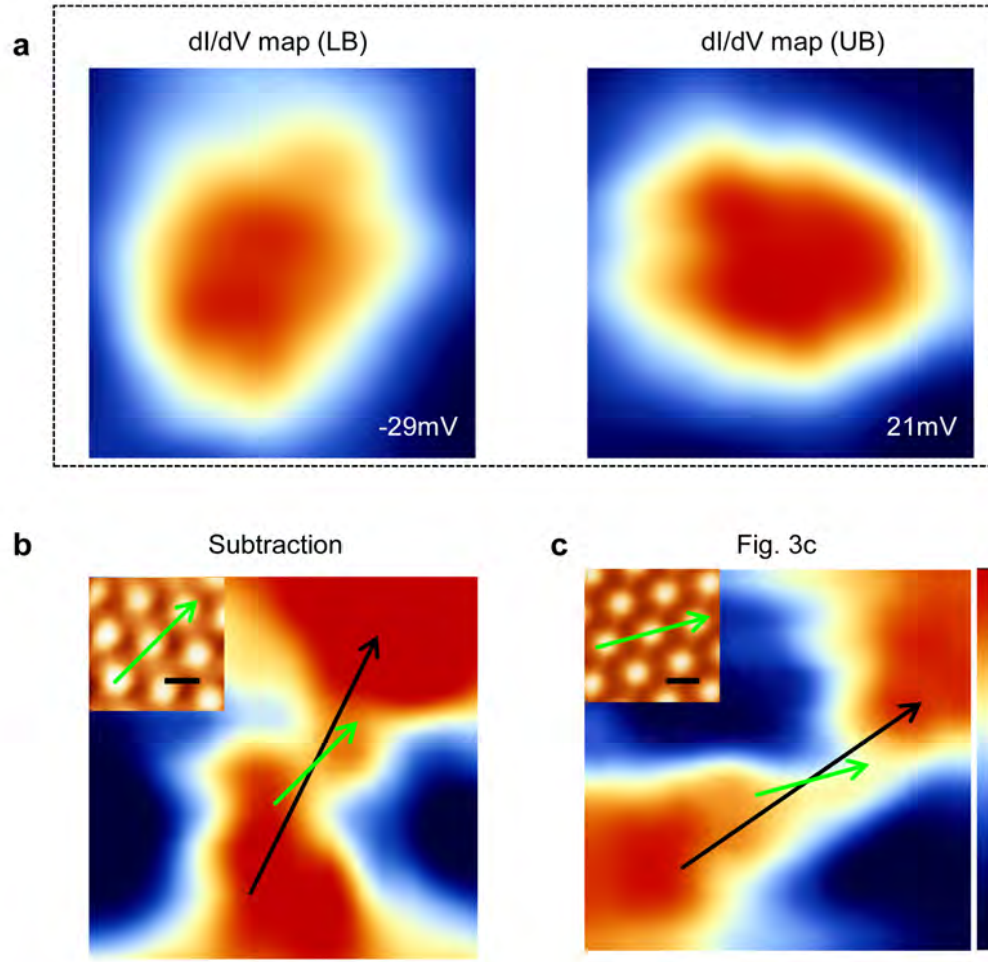

# Extended Data Figure 9

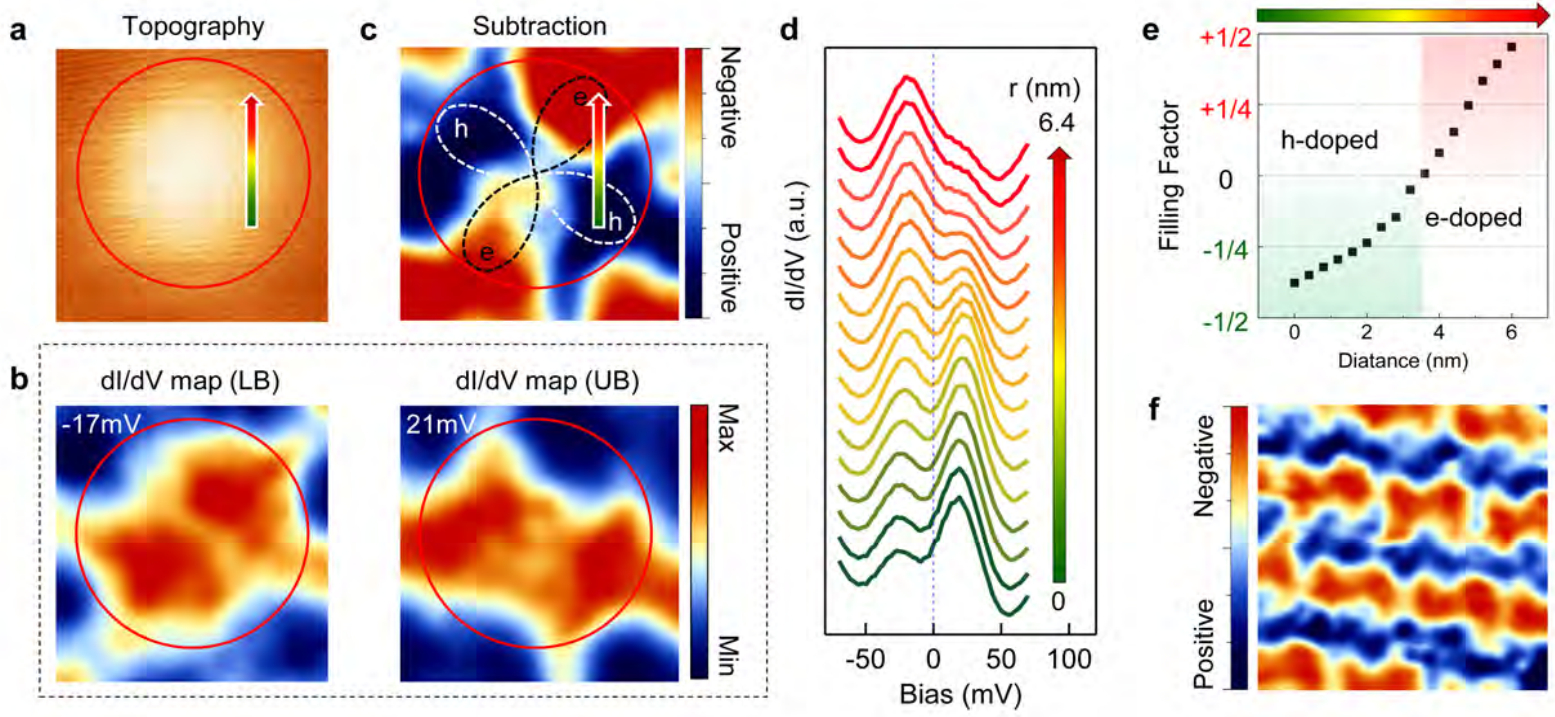

# Extended Data Figure 10

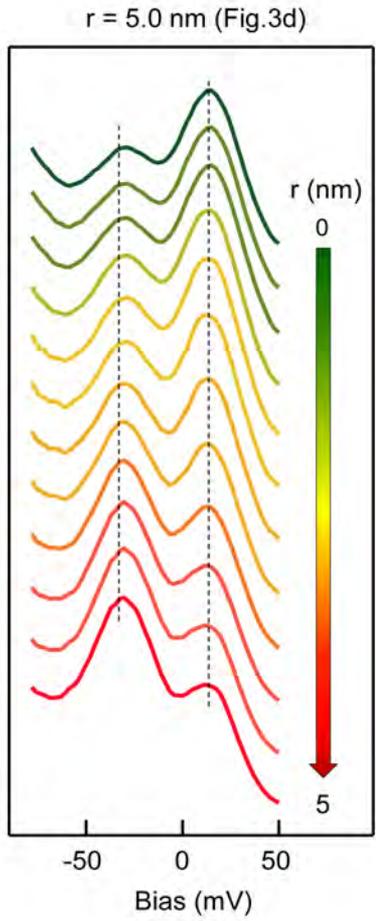
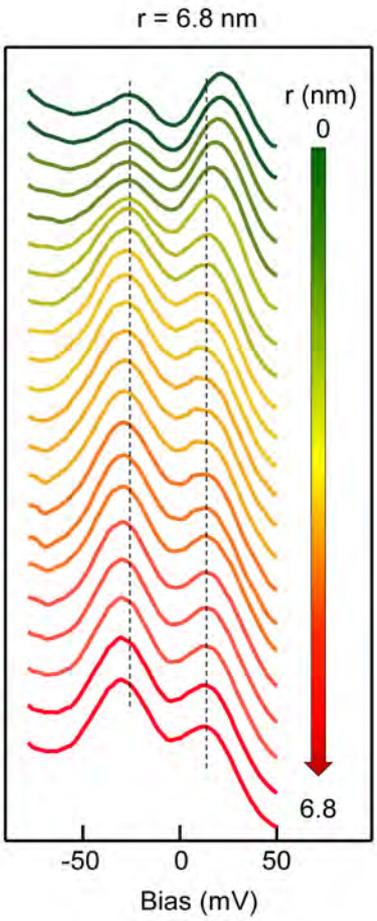